# On the magnetism of the C14 $Nb_{0.975}Fe_{2.025}$ Laves phase compound: Determination of the H-T phase diagram

Maria Bałanda[1] and Stanisław M. Dubiel[2*]

[1]Institute of Nuclear Physics, Polish Academy of Science, PL-31-342 Kraków, Poland
[2]AGH University of Science and Technology, Faculty of Physics and Applied Computer Science, PL-30-059 Kraków, Poland

Abstract

A C14 $Nb_{0.975}Fe_{2.025}$ Laves phase compound was investigated aimed at determining the *H-T* magnetic phase diagram. Magnetization, *M*, and AC magnetic susceptibility measurements were performed. Concerning the former field-cooled and zero-field-cooled *M*-curves were recorded in the temperature range of 2-200K and in applied magnetic field, *H*, up to 1000 Oe, isothermal *M(H)* curves at 2 K, 5 K, 50 K, 80 K and 110 K as well as hysteresis loops at several temperatures over the field range of ± 10 kOe were measured. Regarding the AC susceptibility, $\chi$, both real and imaginary components were registered as a function of increasing temperature in the interval of 2 K - 150 K at the frequencies of the oscillating field, *f,* from 3 Hz up to 999 Hz. An influence of the external DC magnetic field on the temperature dependence of $\chi$ was investigated, too. The measurements clearly demonstrated that the magnetism of the studied sample is weak, itinerant and has a reentrant character. Based on the obtained results a magnetic phase diagram has been constructed in the *H-T* coordinates.

[*] Corresponding author: Stanislaw.Dubiel@fis.agh.edu.pl

## 1. Introduction

The C14 hexagonal $Nb_{1-x}Fe_{2+x}$ compounds known as Laves phases or Frank-Kasper ($\varepsilon$, $\rho$ or $\lambda$) phases exhibit peculiar magnetic properties. Despite they have been studied since three decades no clear-cut picture concerning the magnetism of these compounds has appeared. Based on magnetization and Mössbauer-effect measurements, Shiga and Nakamura concluded that the ground state of $NbFe_2$ is strongly enhanced Pauli paramagnet [1]. Next, Yamada and Sakata reported that this stoichiometric compound is a weak antiferrimagnet (AF) [2]. In turn, Yamada et al., using magnetization and nuclear magnetic resonance techniques, suggested that $NbFe_2$ is AF with ferromagnetic (FM) and AF spin fluctuations [3]. Crook and Cywinski proposed a magnetic phase diagram for the system in the compositional range $-0.04 \leq x \leq 0.04$, according to which the AF phase exists in the close vicinity of $x=0$ and the Néel temperature ranges between ~10 K ($x \approx -0.012$) and ~27K ($x \approx 0.008$), while the magnetic ground state is a mixture of AF and FM phases [4]. The most recent version of the magnetic phase diagram [5] only partially agrees with the one outlined in [4]. Namely, in both phase diagrams the FM ordering exists for Fe- and Nb-over doped compounds, but border $x$-values and those of the Curie temperature, $T_C$, are meaningfully different. Yet the most significant difference occurs for the $x$-range close to zero where, according to Crook and Cywinski, the AF and FM phases coexist, whereas according to the authors of Ref. 5 the two FM phases are separated by a paramagnetic phase with a quantum critical point located at $x \approx -0.015$, and a spin-density waves phase existing between $-0.015 \leq x \leq 0.003$ at $T=0K$. In disagreement with both phase diagrams is a ferrimagnetic ordering suggested recently for a $Nb_{0.985}Fe_{2.015}$ compound based on spin-dependent Compton scattering study supported by *ab initio* electronic structure calculations [6]. It should be also noticed that some features characteristic of a re-entrant spin-glasses (RSG) behavior were observed both for the stoichiometric compound [7] as well as the one over doped with Fe [8].

Motivated by the lack of the clear-cut knowledge and even controversies on magnetism of the $Nb_{1-x}Fe_{2+x}$ compounds we carried out magnetization and AC magnetic susceptibility measurements on a $Nb_{0.975}Fe_{2.025}$ sample as a function of magnetic field and frequency. The obtained therefrom results are presented and discussed in this paper.

## 2. Experimental

The alloy of the nominal composition $NbFe_2$ was produced from Fe (99.95 % pure) and Nb (99.9 % pure) by levitation melting followed by subsequent casting into a pre-heated Cu crucible (15 mm in diameter) with a temperature of 1200°C. The temperature was kept at 1200°C for 45 min. and then the ingot was slowly cooled to room temperature with a rate of 5°C/min. Melting, casting and cooling was performed in an argon atmosphere. Impurity contents of non-metal elements C, N and O in the alloy was found to be less than 100 ppm. The phase identification and characterization of the sample obtained after the applied heat treatment was done by recording XRD patterns. The chemical composition of the final product was determined as $Nb_{0.975}Fe_{2.025}$ by EPMA measurements with a Joel JXA-8100 instrument. Magnetic measurements were performed on a bulk sample, whose mass was 91 mg, in a MPMS XL SQUID magnetometer. Series of zero-field cooled (ZFC) and field-cooled (FC) magnetization temperature dependences, $M_{ZFC}(T)$ and $M_{FC}(T)$, were recorded on heating at several values of the applied magnetic field, $H$. Isothermal magnetization curves, $M(H)$, up to the field of 70 kOe were measured at 2 K, 5 K, 50 K, 80 K and 110 K. Magnetic hysteresis loops were recorded in the temperature interval between 2 K and 50 K over the field range of ± 10 kOe. Both real, $\chi'$, and imaginary, $\chi''$, components of the AC magnetic susceptibility, $\chi = \chi' - i\chi''$, were registered as a function of increasing temperature in the interval of 2 K - 150 K at the frequencies of the oscillating field, $f$, from 3 Hz up to 999 Hz and the amplitude of 2 Oe. An influence of the external DC magnetic field on the temperature dependence of $\chi$ was investigated, too.

## 3. Results and discussion

### 3. 1. Magnetization measurements

#### 3.1.1. M(T) curves

A set of $M_{ZFC}(T)$ and $M_{FC}(T)$ dependences measured in various applied magnetic fields is shown in Fig. 1.

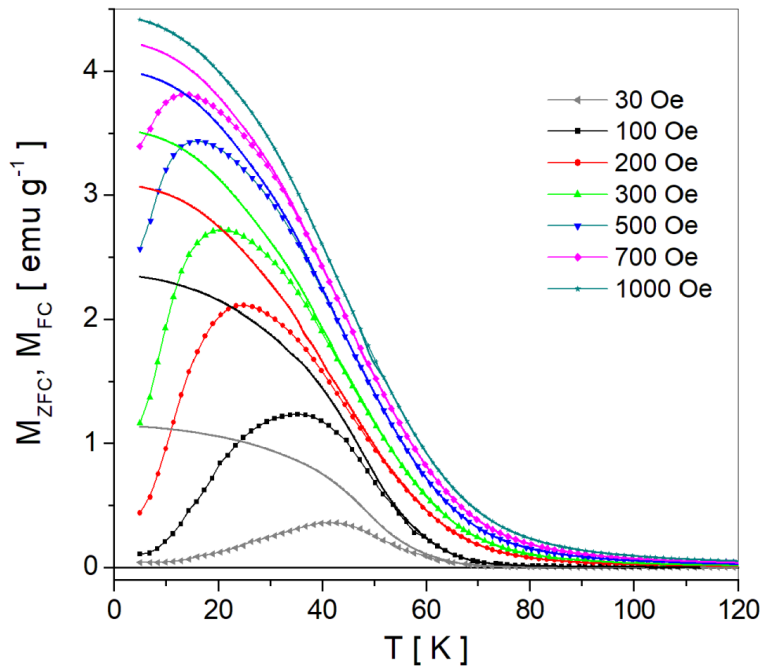

Fig. 1 A set of $M_{FC}(T)$ and $M_{ZFC}(T)$ curves vs. temperature, $T$, recorded in different magnetic fields, $H$, shown in the legend.

The $M(T)$-curves clearly exhibit a bifurcation effect which is no more visible in the curve measured in the field of 1000 Oe. The bifurcation is, among other, one of characteristic features of re-entrant spin-glasses (RSG). Three temperatures relevant to a construction of the magnetic phase diagram in the $H$-$T$ plane can be derived from these measurements viz.

the magnetic ordering temperature (Curie point), $T_C$, the irreversibility (bifurcation) temperature, $T_{ir}$, and the cross-over temperature, $T_{co}$. There are several ways of determining $T_C$, inflection point of the $M_{FC}(T)$ curve and tangent methods are most frequently used ones. Applying the tangent method for the $M_{FC}(T)$ curves shown in Fig.1, we have obtained values of $T_C(H)$ for the field range 30 Oe - 1000 Oe, while the Curie temperature for H ≈ 0, $T_C$ = 63 K, has been determined from the inflection point of AC susceptibility (see § 3.2). $T_{ir}$ is rather ill-defined so its determination is rather arbitrary. In practice for its determination one considers the temperature dependence of the difference between the two magnetization curves i.e. $\Delta M = M_{FC} - M_{ZFC}$ which, for the present case, are displayed in Fig. 2. $T_{ir}$ is the temperature at which $\Delta M$ starts to depart from zero. It signifies the beginning of freezing of the transverse component of the spin and marks a transition from the FM or AF phase to the RSG one. Its dependence on the applied magnetic field is one of two lines characteristic of the behavior of spin-glasses in magnetic field [9].

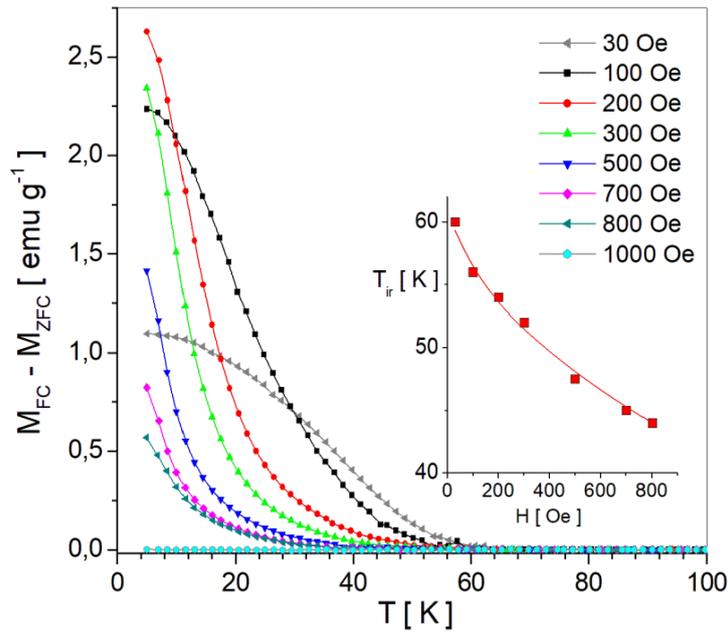

Fig. 2 Difference between the $M_{FC}$ and $M_{ZFC}$ curves vs. temperature, T, for displayed values of the applied magnetic field, H. The inset shows $T_{ir}$ vs. H.

Finally, the temperature $T_{co}$ is the one at which the $M_{ZFC}$ curve has its maximum. It is regarded as the temperature marking a transition, within the RSG state, from a weak into a strong irreversibility regime in which also the longitudinal component of the spin freezes. Its dependence on H defines the so-called Almeida-Touless (AT) line [11]. Both models i.e. GT and AT predict a decrease of $T_{ir}$ and $T_{co}$, respectively, with H which can be expressed by the following formula:

$$T_{ir,co}(H) = T_{ir,co}(0) - aH^\phi \qquad (1)$$

where $\phi$ =2 for GT and $\phi$ =2/3 for AT models. The $T_{ir}(H)$- line is known as the Gabay-Toulouse (GT) line [9] and the $T_{co}(H)$-line can be termed as the de Almeida-Thouless (AT) line [10], because its *H*-dependence is the same as that of the transition line between the weak and the strong irreversibility "phases" of the RSG [9].

It should be also mentioned that for the RSG a linear dependence on *H* i.e. with $\phi$ =1 was predicted [12] and reported to be the case for several systems [12-16].

Based on the values of $T_C(H)$, $T_{ir}(H)$ and $T_{co}(H)$ determined from the *M(T)* curves measured in different magnetic fields, the magnetic phase diagram in the *H-T* plane has been outlined. It is presented and discussed in section 3.3.

The $M_{FC}(T)$ data recorded at *H*=500 Oe were used to determine the DC susceptibility, $\chi_{DC}=M_{FC}/H$. Its inverse dependence on temperature is shown in Fig. 3. The linear part observed for $T \geq \sim 100K$ was fitted to the Curie-Weiss law yielding the Curie-Weiss temperature $\theta$=84K and the effective magnetic moment per formula unit $\mu_{eff}$=1.83 $\mu_B$. The value of $\theta$ can be used to determine the degree of frustration, DOF=$\theta/T_C$=1.33, and the value of $\mu_{eff}$ to determine the degree of itineracy of magnetism in the studied compound, DOI=$q_c/q_s$, following the Rhodes-Wohlfarth criterion [16]. The $q_c$ can be determined from the effective paramagnetic moment $\mu_{eff} = \sqrt{q_c(q_c + 2)}$ $\mu_B$ and $q_s$ from $\mu_s$=2$q_s$ $\mu_B$, $\mu_s$ being the magnetic moment in saturation (here 0.378 [$\mu_B$/f.u.]) [17]. In this way one arrives at DOI=5.74 implying a highly itinerant magnetism in the investigated sample [16].

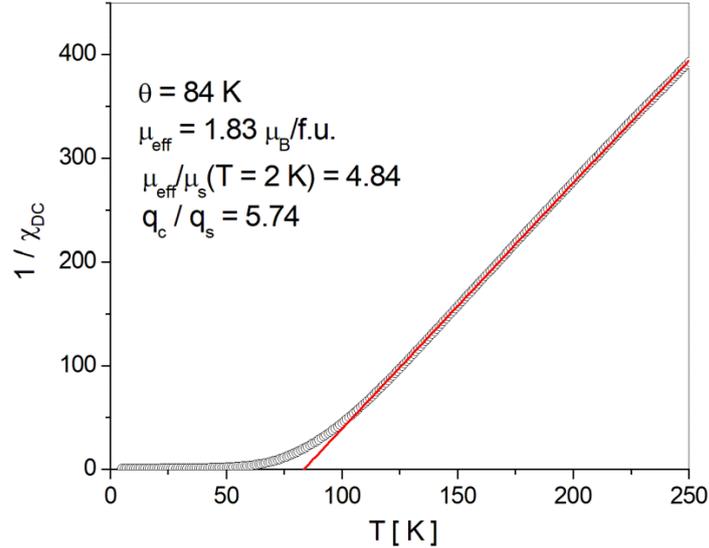

Fig. 3 Inverse DC magnetic susceptibility, $1/\chi_{DC}$, vs. temperature, *T*. The linear part was fitted to the Curie-Weiss law.

### 3.1.2. M(H) curves and hysteresis

Isothermal *M(H)* curves in the field up to 70 kOe are depicted in Fig. 4. Two features should be noticed: (a) sudden increase of *M* in the low field range for 2K and 50K data and (b) visible curvature in the 80K data. The former is in line with the *M(H)* data measured along the easy *c*-axis of the nearly stoichiometric NbFe$_2$ [19] and testifies to a considerable anisotropy visible even in our non-single crystalline sample. The latter was already seen in samples with itinerant magnetism and can be understood in terms of short-range correlations that exist above the Curie temperature e. g. [20,21].

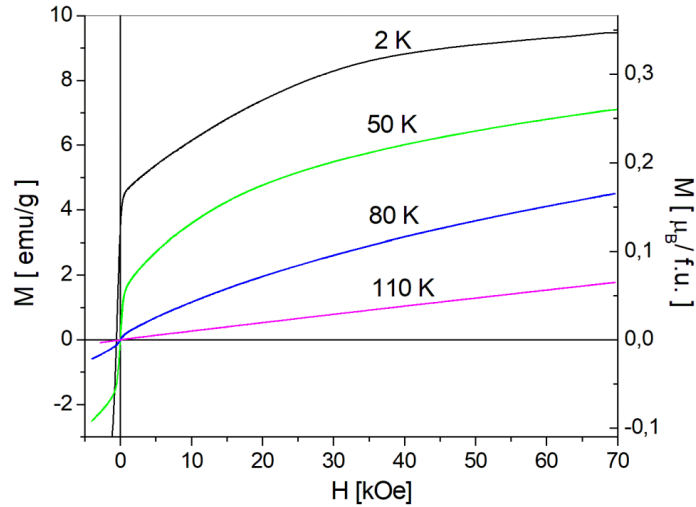

Fig. 4 Isothermal *M(H)* curves labelled with temperatures of measurement. Noteworthy is a step-like increase observed at 2K and 50K curves. Some curvature is seen in the data recorded at 80K i.e. above the Curie temperature.

Hysteresis loops, displayed in Fig. 5, are symmetric and smooth. The temperature dependence of the coercive field, $H_C$, derived from the hysteresis loops, is displayed in the inset of Fig. 5. Its variation shows a monotonic decrease with increasing temperature what, together with the smoothness of the loops themselves, can be taken as evidence in favor of a good sample's chemical homogeneity [5,8]. Initially it is rapid but slows down as $H_C$ approaches the *T* axis tangentially and merges with it near the magnetic ordering temperature, $T_C$=63K. Noteworthy, this behavior resembles that found for the Au-Fe alloy regarded as the canonical SG [18]. The maximum value of $H_C$(2K)=550 Oe is much higher than those reported for x=0.04 ($H_C$=200 Oe at 1.8K) [5] and for x=0.063 ($H_C$=107 Oe at 4.2K) [8]. The difference may have its origin in different composition.

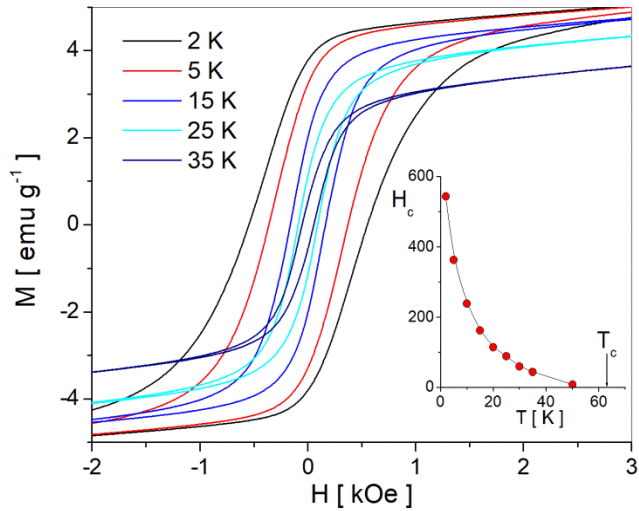

Fig. 5 Set of selected hysteresis loops. The inset shows the dependence of the coercive field, $H_c$, on temperature, $T$. The arrow indicates the position of the Curie temperature, $T_C$.

**3. 2. AC magnetic susceptibility**

AC magnetic susceptibility measurements were performed to further elucidate the nature of magnetism in the $Nb_{0.975}Fe_{2.025}$ sample. The temperature dependence of the real, $\chi'$, and the imaginary, $\chi''$, components of the AC magnetic susceptibility, measured at various frequencies, $f$, between 3 and 999 Hz, are presented in Fig. 6a. As expected, the imaginary component is one order of magnitude smaller than the real one. The shape of the $\chi'(T)$ curves is smooth yet it does not resemble that characteristic of canonical spin glasses e.g. CuMn, AuMn, AuFe for which a well-defined cusp with concave slopes was observed [10]. Instead, the $\chi'$- profile is asymmetric with a steep slope on the low temperature side and a convex one on the high temperature side. The peak-values are rather well-defined but they do not show measurable dependence on frequency – Fig. 6b. Noteworthy, the shape is different than the ones reported previously for Fe-over doped compounds [5,8], where observed irregularities were interpreted as due to in-homogeneities of the samples.

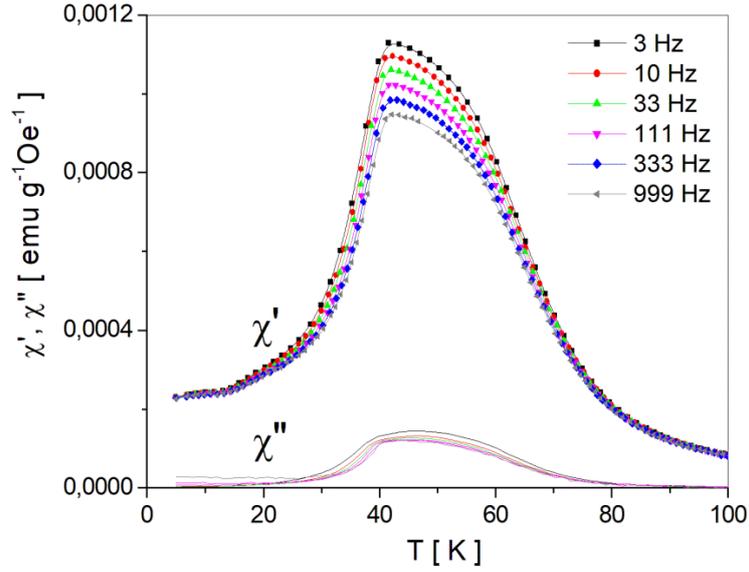

Fig. 6a Real (major), $\chi'$, and imaginary (minor), $\chi''$, components of the AC susceptibility vs. temperature, $T$, measured at various frequencies shown in the legend.

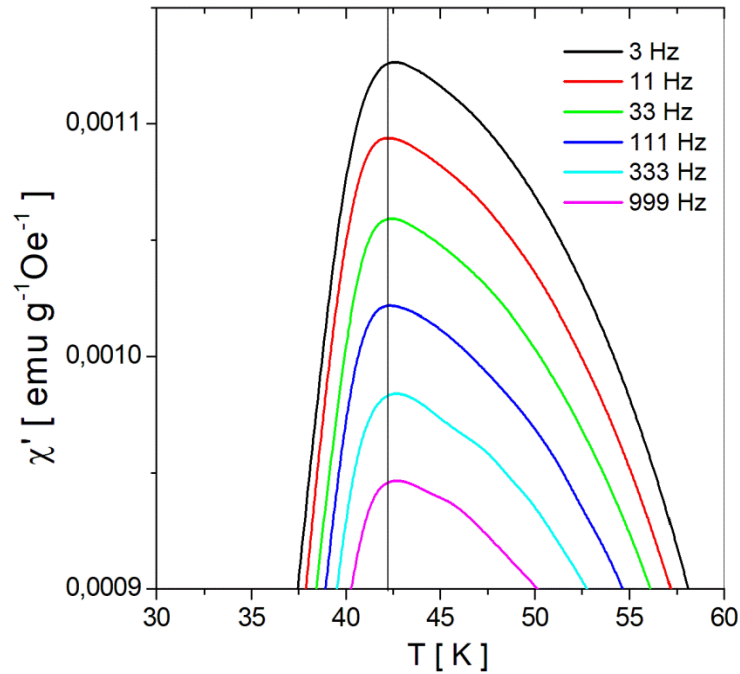

Fig. 6b Profiles of $\chi'$ in the vicinity of their maxima recorded for various frequencies shown. The vertical line marks the position of the maximum for $f$=3 Hz.

From this behavior one could conclude that either the maximum in $\chi'$ at ~42K cannot be associated with the spin-freezing temperature, $T_f$, or the range of the interactions is very long. The latter is very likely in our case as the magnetism of the studied sample is highly itinerant. However, when we have plotted the temperature derivative of $\chi'$ the effect of the frequency is clearly seen – see Fig. 7. Namely, the position of the peak, $T_f$, corresponding to the inflection point of the steep slope shifts with $f$ to higher values, an effect characteristic of spin-glasses. The relative shift of $T_f$ –per decade of frequency, *RST*, can be calculated from the following formula:

$$RST = \frac{\Delta T_f / T_f}{\Delta \log f} \qquad (1)$$

The value of RST is customary used as a figure of merit in order to distinguish between spin-glasses and super paramagnets [22]. They have been also used to differentiate between various types of SGs i.e. canonical and cluster ones. The presently determined value *RST*=0.015 testifies to a weak frequency dependence of $T_f$, and is intermediate between those characteristic for canonical spin glass [23] and cluster spin glass [24]. Noteworthy, similar values of *RST* were found recently for σ-phase Fe-Mo compounds [25] which also exhibit strongly itinerant magnetism with the PM→FM→SG re-entrant character [26] and belong to the same family of the Frank-Kasper phases.

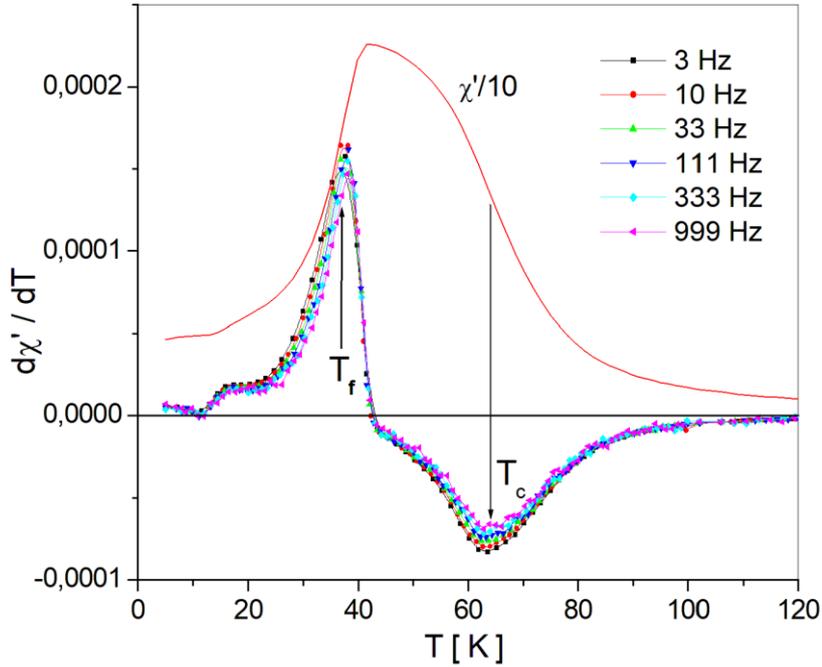

Fig. 7 Temperature dependence of $d\chi'/dT$ for the measured frequencies. The profile of $\chi'$, diminished by a factor of 10, is shown for comparison. The minimum in $d\chi'/dT$ corresponding to the inflection point in $\chi'$ is indicated by an arrow and denoted by $T_C$. Noteworthy, the position of the minima hardly depend on frequency which supports interpretation of $T_C$ as the temperature at which the PM→FM transition occurs. The temperature of the $d\chi'/dT$ maximum is indicated by $T_f$.

To further characterize the spin dynamics, the frequency dependence of the peak in $d\chi'/dT$ was next analyzed in terms of the Vogel-Fulcher law:

$$f = f_0 \exp(-\frac{E_a}{k_B(T_f - T^{VF})}) \quad (2)$$

where $E_a$ is the activation energy, $k_B$ stands for the Boltzmann constant and $T^{VF}$ is the Vogel-Fulcher temperature. The best-fit of equation (2) to the $f$-data is presented in Fig. 8. The parameters yielded from this fit ($E_a/k_B$=132.5 K, $f_0$=6·10$^{11}$Hz, $T_f(f=0)$=36.6 K) can be further used to characterize the degree of clustering and that of the interactions between the spin clusters in the studied sample. The relevant figures of merit introduced by Tholence [25] are defined by $\Theta_1=(T_f-T^{VF})/T_f$ and $\Theta_2=E_a/k_B \cdot T_f$. The value of $\Theta_1$=0.126 is characteristic of metallic spin-glasses [26], and the one of $\Theta_2$=3.6 implies a moderate coupling between the interacting magnetic entities. This, in turn, corroborates with the itinerant character of magnetism in the $Nb_{0.975}Fe_{2.025}$ compound. The value of the pre factor, $f_o$, is one order of magnitude smaller than that typical of a canonical spin-glass system but, at least, one order of magnitude higher than that characteristic of cluster spin-glasses [26].

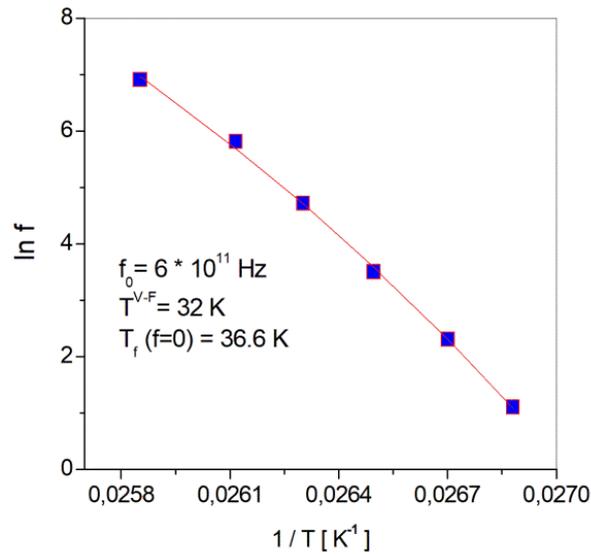

Fig. 8 Relationship between the frequency, $f$, and temperature of the peak position, $T_f$, in the data shown in Fig. 7. The best-fit values of the parameters involved in eq. (2) are displayed.

An effect of the applied magnetic field on $\chi'$ and $\chi''$ curves measured on our sample can be seen in Fig. 9: the apparent decrease of the intensity and smearing of the transition by the field is clear. The data can be further used to determine the effect of the applied magnetic field on the spin-freezing temperature, $T_f$, the frequency dependence of which was successfully obtained by considering the temperature derivative of $\chi'$ – see Fig. 7. The $d\chi'/dT$ curves vs. temperature for different values of $H$ are displayed in Fig. 10. A shift of the maximum defining $T_f$ with $H$ is evident. All the $T_f(H)$ data obtained in this way are presented in Fig. 11.

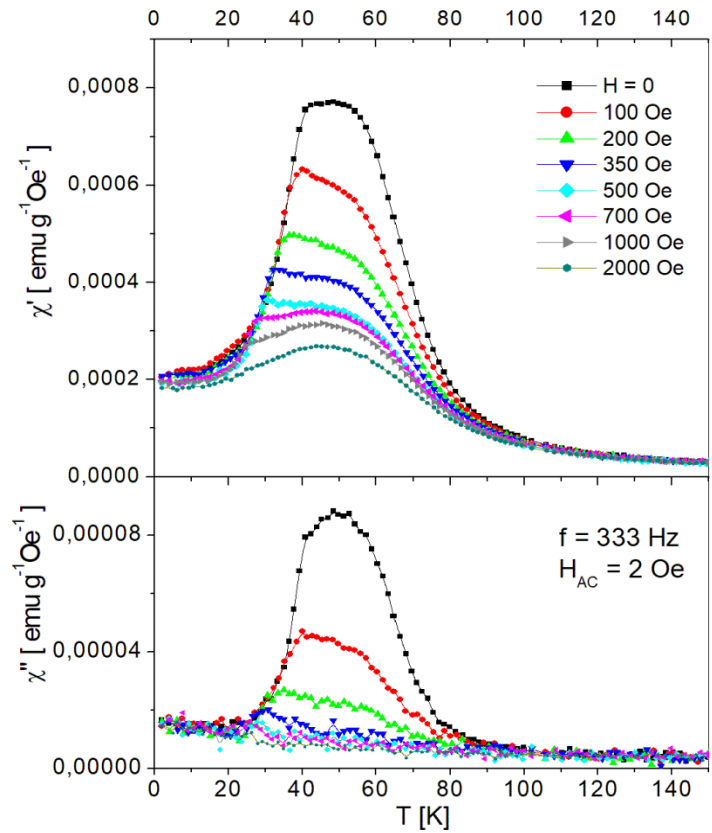

Fig. 9 Real, $\chi'$, and imaginary, $\chi''$, components of the AC susceptibility as measured under conditions described in the legend.

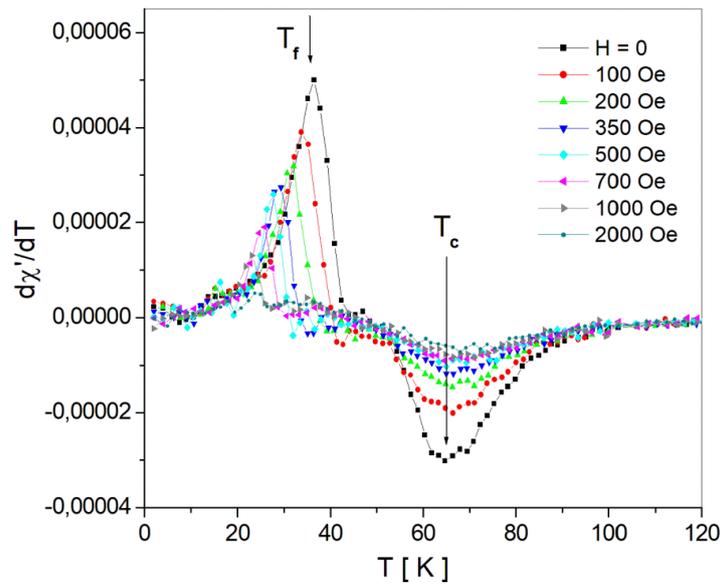

Fig. 10 Temperature dependences of $d\chi'/dT$ measured in different magnetic fields shown. The spin-freezing temperature, $T_f$, and the Curie temperature, $T_C$, are indicated.

## 3. 3. H-T phase diagram

Based on the measurements described above and taking into account the obtained results the magnetic phase diagram in the *H-T* coordinates has been constructed. The one which emerges from the present study can be seen in Fig. 11.

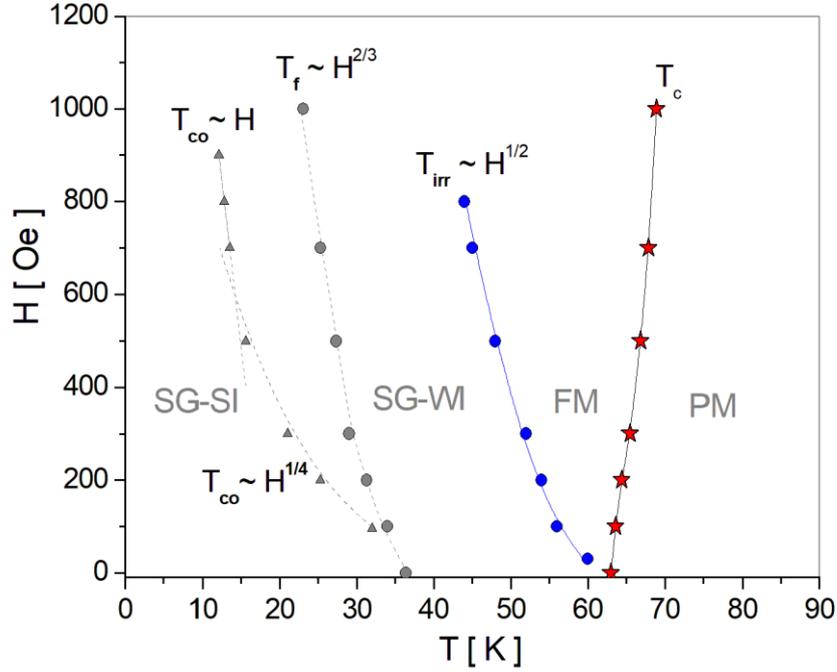

Fig. 11 *H-T* magnetic phase diagram of the investigated sample constructed using the results obtained in the present study. The lines represent fits to the data in terms of eq. (1).

Our results give clear-cut evidence that the magnetism of the $Nb_{0.975}Fe_{2.025}$ C14 Laves phase compound has a reentrant character i.e. on lowering temperature one observes the following sequence of transitions: PM→FM→SG. This finding does not agree with the most recent phase diagram proposed by Moroni-Klementowicz and co-workers [5] according to which for the sample with x=0.025 there is only PM →FM transition with $T_C$≈47K. On the other hand, the phase diagram outlined by Crook and Cywinski [4] agrees only qualitatively with ours as in their phase diagram the mixed state (FM+AF), which corresponds to the presently found RSG, terminates already at *x*≈0.015 and the FM phase exists for higher *x*-values. There is also a disagreement concerning the value of the Curie temperature, $T_C$, at x=0.025: for our sample $T_C$≈63K while the corresponding value reported in Ref. 4 is ~55K. Magnetization and AC magnetic susceptibility measurements performed in the applied magnetic field enabled construction of the magnetic phase diagram in the *H-T* coordinates.

Its structure, as shown in Fig. 11, can be described in terms of lines defined by the characteristic temperatures: (1) the Curie temperature, $T_C$, (2) the irreversibility or bifurcation temperature, $T_{ir}$, (3) the spin-freezing temperature, $T_f$, and (4) the cross-over temperature, $T_{co}$. The *H*-dependence of $T_{ir}$ is theoretically related to the so-called GT-line for which $\phi=2$. However, as indicated in Fig. 11, the presently found $T_{ir}(H)$-dependence is very different, namely $\phi=0.5$. The fragment of the phase diagram confined between $T_c$ and $T_{ir}$ represents the FM regime. It is evident that its phase range is very narrow at H≈0. However, the applied magnetic field strengthens the FM phase which manifests itself in a weak increase of the Curie temperature and the decrease of $T_{ir}$. Consequently, the range of the FM phase increases with *H*. The $T_{ir}(H)$-line constitutes the upper border of the SG-regime with a weak irreversibility marked by SG-WI in the phase diagram. Its lower border and simultaneously the upper border of the SG-regime with a strong irreversibility, labelled as SG-SI, is defined by $T_{co}(H)$ or $T_f(H)$ lines which can be termed as the "AT-line". In the present case the two lines do not overlap and show different *H*-dependence. Whereas the $T_f(H)$ is in line with the expectation i.e. it follows the $H^{2/3}$ law, the $T_{co}(H)$ is not. It shows a crossover from a non-linear behavior with $\phi=0.25$ at low *H*, to a linear behavior at higher *H*. Noteworthy, the crossover from the non-linear to the linear behavior in both lines i.e. $T_{ir}(H)$ and $T_{co}(H)$ was recently revealed in σ-phase Fe-Cr and Fe-V compounds [15,30]. It should be emphasized that the AT-line is regarded as one of the most important features of spin-glasses [31]. However, our present study shows that there is no unique way of determining this line, at least for the studied compound. The two lines we have determined – one from the AC susceptibility measurements and the other one from the DC magnetization curves measured in the ZFC-protocol - do not overlap with each other and have different dependence on *H*. These observations cast some doubt on the applicability of the mean-field model [9]. In any case, the magnetism of the $Nb_{0.975}Fe_{2.025}$ C14 Laves phase is more complex than it follows from the known phase diagrams [4,5] as it exhibits a reentrant i.e. PM→FM→SG behavior which was not reported so far.

## 4. Conclusions

The in-field DC and AC magnetic susceptibilities study on the C14 $Nb_{0.975}Fe_{2.025}$ Laves phase compound enabled construction of the magnetic phase diagram in the *H-T* coordinates. The results obtained can be concluded as follows:

1. The magnetism of the studied sample is weak, itinerant and has a re-entrant character i.e. the PM→FM→SG sequence of transitions is takes place.

2. The temperature of the FM-ordering $T_C$≈63K at *H*=0, and it weakly increases with *H* reaching the value of ~68K at *H*=1000 Oe.

3. The magnetic ground state is constituted by a SG which is heterogeneous i.e. consists of a sub-phase with a weak irreversibility, SG-WI, and of the one with a strong irreversibility, SG-SI.

4. The border between the FM and the SG-WI phases shows a $H^{1/2}$-dependence i.e. does not agree with the Gabay-Toulouse prediction.

5. The border between the SG-WI and the SG-SI regimes is not uniquely defined: if one considers the $T_f$ temperature than the corresponding line follows the expected $H^{2/3}$-law, but if one takes into account the $T_{co}$ temperature the $T_{co}(H)$-line consists of two parts: the low-field one proportional to $H^{1/4}$, and the high-field one proportional to $H$.

**Acknowledgement**